\documentclass[%
 reprint,
superscriptaddress,
aip,
]{revtex4-1}

\usepackage{xcolor}
\usepackage{lineno}
\usepackage{graphicx}
\usepackage{hyperref}
\usepackage{amsmath}
\usepackage{color}
\usepackage{graphics}
\usepackage{adjustbox}
\definecolor{red}{rgb}{0.8, 0.0, 0.0}
\definecolor{blue}{rgb}{0.06, 0.2, 0.65}
\definecolor{green}{rgb}{0,0.6,0}

\begin{document}

\title{Plastic 
deformation mechanisms during nanoindentation of W, Mo, V body-centered cubic single crystals and their corresponding W-Mo, W-V equiatomic random solid solutions}

\author{F. J. Dominguez-Gutierrez}
\affiliation{%
NOMATEN Centre of Excellence, National Center for Nuclear Research, 
05-400 Swierk/Otwock, Poland
 }%

\author{S. Papanikolaou}
\affiliation{%
NOMATEN Centre of Excellence, National Center for Nuclear Research, 
05-400 Swierk/Otwock, Poland
 }%

\author{S. Bonfanti}
\affiliation{%
NOMATEN Centre of Excellence, National Center for Nuclear Research, 
05-400 Swierk/Otwock, Poland
 }%

\author{M. J. Alava}
\affiliation{%
NOMATEN Centre of Excellence, National Center for Nuclear Research, 
05-400 Swierk/Otwock, Poland
 }%
\affiliation{Department of Applied Physics, Aalto University, P.O. Box 11000, 00076 Aalto, Espoo, Finland}

\vspace{10pt}

\begin{abstract}
 
Deformation plasticity mechanisms in alloys and compounds may unveil the material capacity towards optimal mechanical properties.  We conduct a series of molecular dynamics (MD)
simulations to investigate plasticity mechanisms due to  
nanoindentation in pure tungsten, molybdenum and vanadium body-centered cubic single crystals, as well as the also body-centered cubic, equiatomic, random solid solutions (RSS) of tungsten--molybdenum and 
tungsten--vanadium alloys. 
Our analysis focuses on a thorough, side-by-side comparison of dynamic deformation processes, defect nucleation, and evolution, along with corresponding stress--strain curves. We also check the surface 
morphology of indented samples through atomic shear strain mapping. As expected, the presence of Mo and V atoms in W matrices introduces lattice strain and distortion, increasing material 
resistance to deformation and slowing down dislocation 
mobility of dislocation loops with a Burgers vector
of 1/2 $\langle 111 \rangle$. Our side-by-side comparison displays a remarkable suppression of the plastic zone size in equiatomic W--V RSS, but not in equiatomic W--Mo RSS alloys, displaying a clear prediction for optimal hardening response equiatomic W--V RSS alloys. If the small-depth nanoindentation plastic response is indicative of overall mechanical performance, it is possible to conceive a novel MD-based pathway towards material design for mechanical applications in complex, multi-component alloys.

\end{abstract}

%
\vspace{2pc}
\keywords{
W--Mo alloy, W--V alloy, nanoindentation, plasticity, body-centered cubic, random solid solutions, tungsten, molybdenum, vanadium}
%
%
\maketitle
%
%


Tungsten--Molybdenum (W--Mo)
alloys are promising materials with superior physical 
properties that make them suitable for sustaining extreme operating 
conditions without degradation 
\cite{KRAMYNIN2022106814,CHEN2021105668,LAN2022112140,SAHOO2015124,liu2013nanostructured}.
 As the W percentage increases, the thermal conductivity decreases while 
the strength at both room and high temperatures
increase. 
\cite{CHEN2020153760,TAO2022356,JIANG201856,OHSERWIEDEMANN201327}.
In addition, Mo and W form isomorphic solid solutions due
to their complete solid and liquid solubility, making 
it possible to create solid solutions across the
composition range 
\cite{LAN2022112140,GAFFIN2022105778,WEI2023154534}. 
Nevertheless, it is also known that tungsten exhibits a relatively low 
fracture toughness at low temperatures, with its 
susceptibility to cracking 
\cite{gumbsch1998controlling,ARSHAD201496}. 
As an alternative, the addition of vanadium to W matrices 
(W--V alloys) may enhance its mechanical properties and 
improve creep resistance at elevated temperature 
\cite{PhysRevB.84.104115,JIANG2019165,WEI2023154534,WURSTER2011166}. 
Furthermore, Vanadium has been recently labeled as a unique element 
for strengthening in high entropy alloys and more generically, 
random alloy solutions~\cite{yin2020vanadium}. In this work, we 
investigate the effects of Vanadium in W-based alloys, and we arrive 
to similar conclusion as in Ref.~\onlinecite{yin2020vanadium}.

In this work, we aim to gain insights into the alloys' 
behavior under external loads and the possibility to develop novel material design methods, by investigating 
nanoindentation--induced plasticity through load--displacement
curves and surface patterning  
\cite{oliver_pharr_1992,SCHUH200632,PhysRevLett.109.075502,YU2020135,KARIMI2023115559,FRYDRYCH2023104644}. 
Molecular dynamics (MD) simulations, particularly large-scale 
atomistic simulations, offer a cost--effective alternative
for comprehending plastic deformation mechanisms in W 
and its alloys 
\cite{PITTS2013S48,JavVarilla,KURPASKA2022110639,Javier2021,PhysRevMaterials.7.043603}. The transition from elasticity to 
plasticity is characterized by pop--in events, sudden 
displacements indicating the initiation of dislocation 
sources in small indentation zones, as discussed in our
work 
\cite{Pathak2014,PATHAK20151,PhysRevMaterials.7.043603,KARIMI2023115559}. 
This phenomenon showcases a distinct drop in stress, 
a focal point in our study of equiatomic W-Mo and W-V alloys' 
hardening properties compared to single elements 
\cite{pohl2019popin,LI20111147,XIONG2016214,MULEWSKA202355,naghdi2022dynamic}, despite maintaining the body-centered cubic (BCC) crystalline structure.


We perform computer simulations 
by using the Large--scale Atomic/Molecular Massively Parallel
Simulator (LAMMPS) \cite{THOMPSON2022108171} and
the Embedded atom model with Finis--Sinclair (EAM--FS) 
potential developed for W--Mo and W--V 
by Y. Chen et al. \cite{CHEN2020152020}. 
This potential accurately models various physical and
mechanical properties, $\langle 111 \rangle$ dumbbell 
migration, stacking fault energies, and relative stability
of the $\langle 100 \rangle$ and 1/2$\langle 111 \rangle$ 
interstitial dislocation loops. 
While, the surface energies of these two metals are reported to 
be slightly lower than the DFT results, 
To ensure accuracy in capturing plastic deformation mechanisms during nanoindentation, in Table  \ref{tab:elastic} 
we report calculations for the elastic 
constants (C$_{ij}$), bulk modulus (B = $(C_{11} + 2C_{12})/3)$) 
in GPa, 
shear modulus (G = $(C_{11} - C_{12})/2)$) in 
GPa, poisson's ratio ($\nu$ = C$_{12}$/(C$_{11}$+C$_{12}$)), 
 and the Young's modulus ($E_Y$ in GPa) of 8000 atoms (001) samples
 in single element samples of W, Mo, and V, as well as 
 random equatomic samples of W--Mo and W--V
 alloys.
We report results that are in good agreement with  prior 
calculations, by some of us, for pure crystalline W and Mo~\cite{DOMINGUEZGUTIERREZ202238,PhysRevMaterials.7.043603} 
and those reported in the literature \cite{YU2020135,CHEN2020152020}.

\begin{table}[t!]
    \centering
    %
    \caption{Computed values of elastic constants, shear modulus G 
    (GPa), Bulk modulus B (GPa), Poisson ratio $\nu$, and 
   Young's modulus $E_Y$ (GPa). 
   Compared to DFT calculations \cite{JIANG201856,JIANG2019165}.}
    \resizebox{\columnwidth}{!}{%
    \begin{tabular}{crrrrrrrrrrr}
    \hline \hline
         & W     & DFT& Mo    & DFT & V    & DFT & WMo  & DFT & WV & DFT \\
        \hline
        C$_{11}$ & 522   & 530 & 479  & 472 & 234  & 283 & 507  & 503 & 328 & 396 \\
        C$_{12}$ & 204   & 211 & 162  & 171 & 115  & 143 & 196  & 190 & 142 & 185 \\
        C$_{44}$ & 161   & 141 & 113  &  93 & 45   & 15 & 129  & 115 & 115 & 57 \\
        G        & 159   & 148 & 158  & 116 & 57   & 37 & 155  & 132 & 93 & 76 \\
        B        & 310   & 317 & 268  & 271 & 155  & 190 & 300 & 294 & 204 & 255 \\
        E        &  407  & 384 & 397  & 304 & 157  & 105 & 398  & 344 & 243 & 208 \\
        $\nu$    &  0.28 & 0.29& 0.25 & 0.31& 0.33 & 0.4 & 0.28 & 0.30& 0.30 & 0.36 \\
        \hline
    \end{tabular}}
    \label{tab:elastic}
\end{table}

\begin{figure}[b!]
    \centering
    \includegraphics[width=0.48\textwidth]{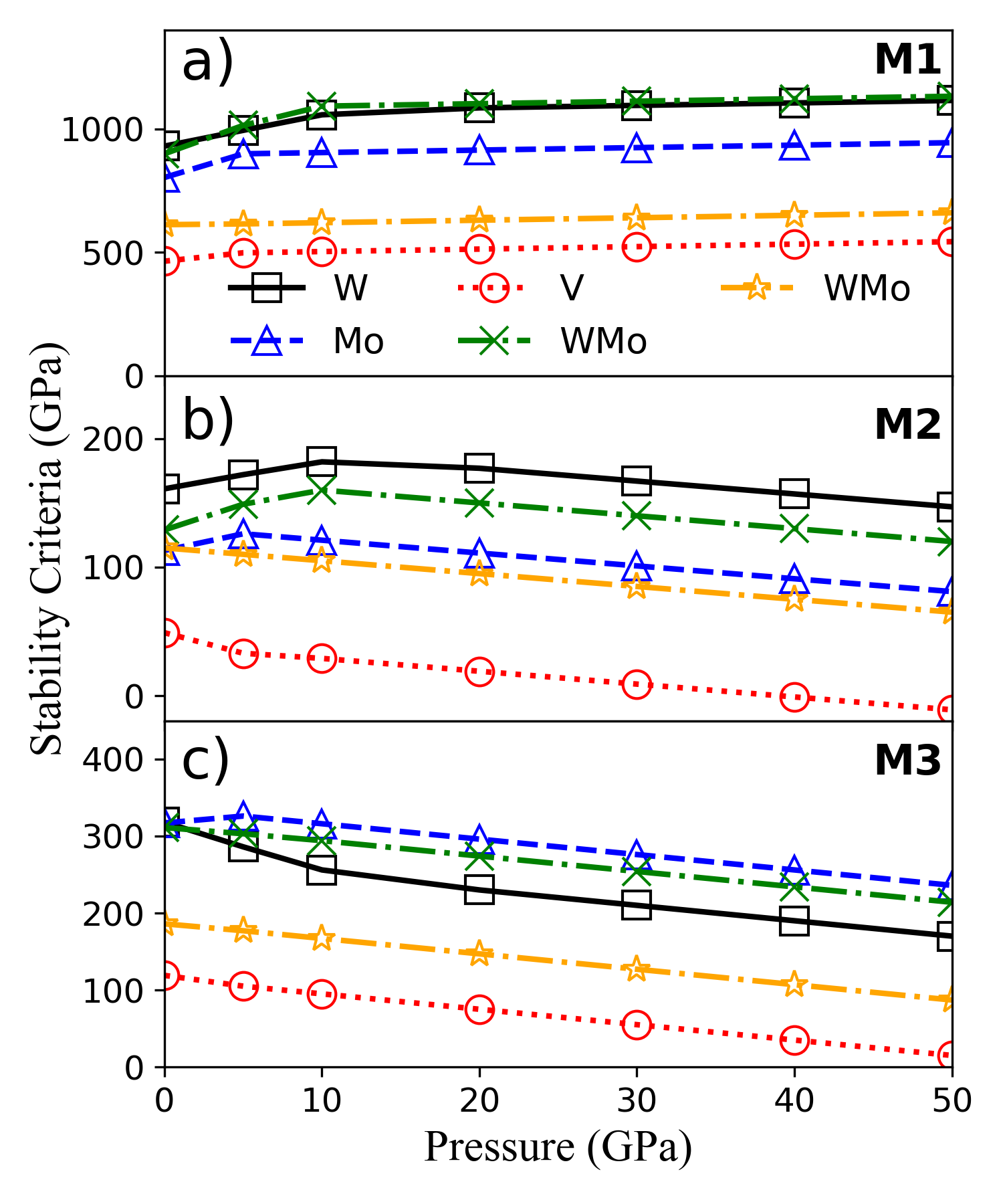}
    \caption{Spinodal, shear, and Born stability criteria with
    hydrostatic pressure for single W, Mo, V, and binary alloy
    WMo and WV. The pressure range showing stability of 
    the interatomic potentials is considered to setup 
    the numerical conditions in the MD simulations.}
    \label{fig:stability}
\end{figure}

To validate the suitability of the EAM--FS potentials 
for simulations under high stress 
conditions, we assess the elastic constants within a 
pressure range relevant to the loading process \cite{XIONG2016214}. 
We further examine the elastic stability conditions by
evaluating the spinodal, shear, and Born criteria 
\cite{Xiong_2014} under hydrostatic pressure $P$. 
These criteria, denoted as M1 = C$_{11}$+2C$_{12} + P$ $> 0$,
M2 = C$_{44}$-$P > 0$, and M3 = C$_{11}$-C$_{12}$-2$P > 0$,
are shown in Fig. \ref{fig:stability}. 
Notably, the materials exhibit stability from 0 to 50 GPa,
which is taken into account to set up the numerical 
environment for nanoindentation simulations.

For modeling nanoindentation, the initial BCC samples of W, Mo, 
and V were created with lattice constants of approximately 
0.316 nm, 0.315 nm, and 0.303 nm, 
respectively. The dimensions of the samples were chosen to be 
approximately 50 nm × 51 nm × 52 nm for the [001] orientation, 
containing 8,190,720 atoms, and 8,115,360 atoms for the [011] 
samples. 
For the [111] orientation, the sample size was approximately 
50 nm × 51 nm × 52 nm, containing 8,424,900 atoms.
The $x$ and $y$ axes have periodic boundary 
conditions to simulate an infinite surface, while 
the $z$ orientation has a frozen bottom boundary for
numerical stability, a thermostatic section above the
frozen layer to dissipate the heat generated during
nanoindentation, and a ``dynamical" section for
atoms--indenter interaction. 
Additionally, a 5~nm vacuum region is added atop the 
sample as an open 
boundary~\cite{KURPASKA2022110639,DOMINGUEZGUTIERREZ2021141912}.
The W--Mo and W--V alloys are created by random 
substitution of 50\% of W atoms from the initial BCC W sample 
with Mo or V atoms, as done previously, an 
assumption supported by thermodynamic binary alloy phase diagram 
calculations~
\cite{turchi2005application,muzyk2011phase}. 
The FIRE (fast inertial relaxation engine) 2.0 protocol 
\cite{GUENOLE2020109584} is then used to optimize 
the energy of the samples towards the lowest energy structure. 
Then, we equilibrate the samples for 100~ps using a 
Langevin thermostat at 300~K with a time constant of 100~fs 
\cite{DOMINGUEZGUTIERREZ2021141912}. 

\begin{figure*}[t!]
    \centering
    \includegraphics[width=0.48\textwidth]{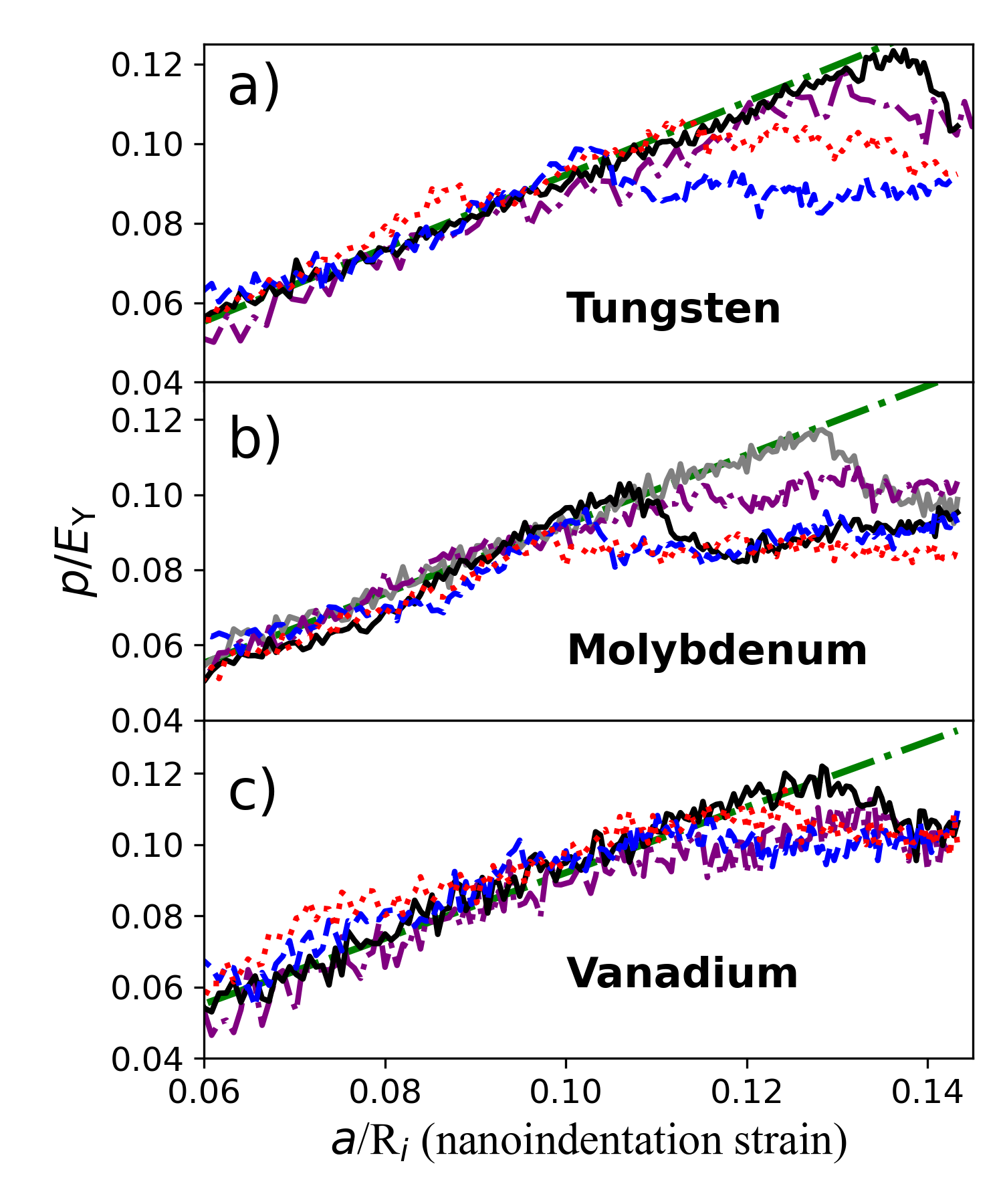}
    \includegraphics[width=0.48\textwidth]{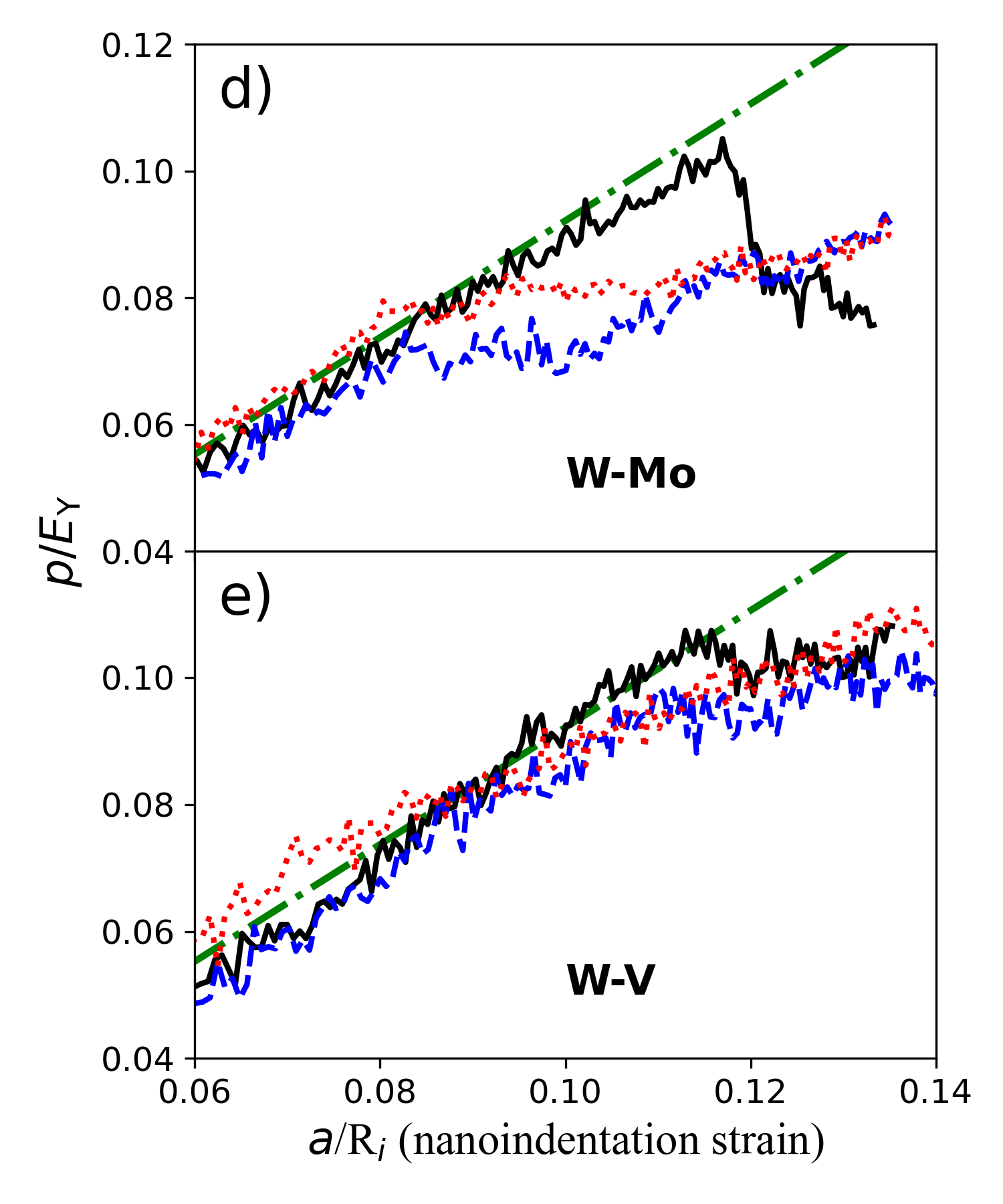}
    \caption{Average contact pressure evolution, $p$, 
    normalized by Young’s modulus, is depicted with normalized 
    contact radius for W, Mo, and V matrices, along with W--Mo
    and W--V alloys. The figure employs a color scheme: 
    solid black line for [001] orientation, dashed blue line
    for [011], and dotted red line for [111]. The results
    conform to the universal linear relationship
    $0.844/(1-\nu^2$ )$a/R_i$, illustrated with the 
    green dashed--dotted line. To validate our findings,
    comparisons with tabGAP 
    \cite{JesperTabGAP,PhysRevMaterials.7.043603}, depicted
    with purple double dotted--dashed lines, and EAM-AT+ZBL 
    \cite{ESalonen_2003,DOMINGUEZGUTIERREZ2021141912} simulations, 
    represented by gray solid lines, are included.}
    \label{fig:single_samples}
\end{figure*}

The indenter tip is treated as a non--atomic repulsive 
imaginary (RI) rigid sphere, with a force potential 
defined as $F(t) = K \left(\vec r(t) - R_i \right)^2$, 
where $K = 236$ eV/\AA$^3$ (37.8 GPa) represents the
force constant, and $\vec r(t) = x_0 \hat x + y_0 \hat y 
+ (z_0 \pm vt)\hat z$ denotes the center's 
position of the tip over time with $x_0$ and $y_0$ as 
the center of the surface sample on the $xy$ plane, 
and $z_0 = 0.5$ nm is the initial gap between the 
surface and the indenter tip. Given the restricted 
surface information in EAM potentials as observed in 
our previous work \cite{PhysRevMaterials.7.043603}, 
a tip radius $R_i = 15$ nm is chosen to accurately 
model the elastic--to--plastic deformation transition, 
and a speed of $v = 20$. The tip's center is randomly 
shifted to ten different 
positions to account for statistical variation. 
We use a $NVE$ statistical thermodynamic ensemble with
the velocity Verlet algorithm for 125~ps with a time 
step of $\Delta t = 1$~fs. A maximum indentation depth
of 2.0~nm is selected to minimize boundary layer effects
in the dynamic atoms region.
To identify defects in indented samples, 
we apply the BCC Defect Analysis (BDA) developed by M\"oller
and Bitzek \cite{moller2016bda} which utilizes coordination
number (CN), centrosymmetry 
parameter (CSP), and common neighbor analysis (CNA) 
techniques to detect typical defects found in
bcc crystals. 

\begin{figure}[t!]
    \centering
    \includegraphics[width=0.48\textwidth]{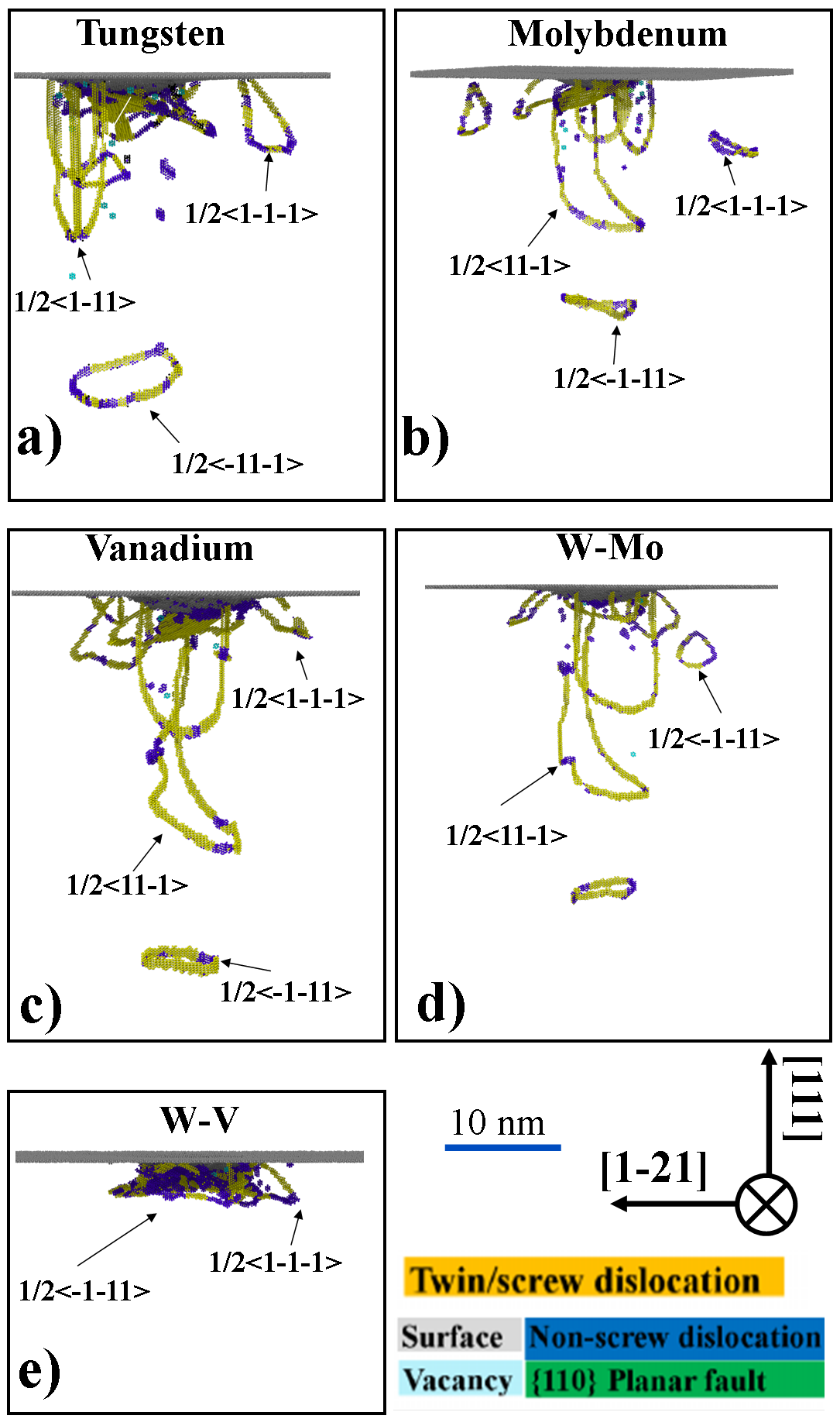}
    \caption{Identification of defects in the plastic region beneath the 
    indenter tip at the maximum depth of the single element tungsten (W), 
    molybdenum (Mo), and vanadium (V) samples, as well as the binary 
    alloys WMo and WV, using the BCC defect analysis (BDA) technique. 
    Screw dislocation/twinning planes are represented by yellow-colored 
    atoms, while blue-colored atoms indicate edge dislocations. The top 
    layer atoms are depicted in gray, and atoms in close proximity to 
    vacancies are illustrated by light blue spheres.}
    \label{fig:WMOVstructure}
\end{figure}

The common pathway to investigate nanoindentation is through average force-displacement curves. Figure~\ref{fig:single_samples} shows 
the ratio between the mean contact pressure~$p$ and Young 
modulus~$E_{\rm Y}$, taken from Tab \ref{tab:elastic} for 
each sample, as a function of the normalized contact 
radius~$\mathrm{a}/R_i$ between the sample and the tip.
The ratio is calculated using a linear elastic contact mechanics 
formulation 
\cite{JavVarilla,REMINGTON2014378}:
\begin{equation}
    \frac{p}{E_Y} = \frac{2\pi}{3E_Y} 
    \left[ 24P_{\rm ave} \left( \frac{E_Y R_i}{1-\nu^2} \right)^2 \right]^{1/3},
\end{equation}
with the average load as 
$P_{\rm ave}=1/N \sum_i^N P_i$ with $N=10$ the number 
of MD simulations; 
and the normalized contact radius is obtained with the 
geometrical relationship \cite{PhysRevMaterials.7.043603}: 
\begin{equation}
\mathrm{a}(h) = \left[ 3PR \left( 
 1-\nu^2\right)/8E_{\rm Y} \right]^{1/3}.
 \end{equation}
 This quantity provides an intrinsic measure of the surface 
 resistance to defect nucleation~\cite{VARILLAS2017431,PhysRevMaterials.7.043603}.
The results demonstrate a universal linear relationship,
as reported in \cite{JavVarilla}, given by $0.844/(1-\nu^2)a/R_i$. 
This is utilized to identify the pop--in event, where the 
contact pressure deviates from the linear 
geometric fitting showing plastic instability initiated as 
a notable pressure drop.


From our MD simulations, the [001] orientation displayed 
the highest critical load, suggesting enhanced resistance
to plastic deformation. Fig. \ref{fig:single_samples}a--c)
compares our outcomes with those from tabGAP 
\cite{byggmastar_simple_2022, JesperTabGAP}, 
used for pure W, Mo, and V \cite{PhysRevMaterials.7.043603},
and EAM-based potentials with ZBL corrections
\cite{ESalonen_2003}, employed for pure Mo as in our
prior work \cite{DOMINGUEZGUTIERREZ2021141912}. 
While tabGAP predictions aligned well for pure W and V
with our EAM-FS findings, the
pressure drop observed in Mo in both EAM-AT+ZBL and EAM-FS 
simulations was a challenge.
Transitioning to the effects on Mo--W and M--V alloys, Fig. \ref{fig:single_samples}d) and e) demonstrates the
influence of Mo and V atoms in the W matrix. 
For Mo, the yield point at (001) is at 0.12 $a/R_i$, 
between those of pure W and Mo, consistent across
orientations. Also, the pressure drop is more pronounced
for [001]W--Mo compared to pure W and Mo. Conversely, V 
atoms in W lead to decreased yield points across 
orientations, in agreement with the V sample but 
distinct from W matrices. Notably, the yield point 
for [001]W--V is 0.11 $a/R_i$, lower than in pure W and V, signifying the onset of exceptional work hardening.

Fig. \ref{fig:WMOVstructure} shows material structures of 
selected MD simulation at maximum indentation depth for
the [111] orientation, determined by the BDA
method. Single--element matrices form a single
dislocation loop on the \{111\} slip plane, along with an
embryonic dislocation (Burgers vector 1/2$\langle1-1-1\rangle$) 
traveling the \{112\} plane. A twinning plane also forms
near the indenter, consistent with prior W simulations 
\cite{PhysRevMaterials.7.043603}. Mo initiates a single 
dislocation loop and creates an embryonic
dislocation on the \{112\} plane with the same Burgers
vector.
V matrices exhibit shear loop formation under the indenter, 
indicating plasticity. However, embryonic dislocation movement
is limited, favoring lateral motion instead. 
This leads to a prismatic dislocation loop (Burgers
vector 1/2$\langle-1-11\rangle$) followed by a second
loop, typical in BCC materials
\cite{PhysRevLett.109.075502}.
50\% Mo concentration minimally impacts dislocation
nucleation and lateral deformation in W samples. 
However, V atoms significantly alter plastic deformation,
generating shear half--loops and hindering screw
dislocation evolution on \{111\}. [111]W--V alloy's 
structure exhibits crystallographic dependency on plastic 
deformation, displaying ductile behavior. Alongside 
surface twin boundaries fostering screw dislocation
formation, co-planar dislocation arrays 
(1/2$\langle 111 \rangle$ and 1/2$\langle -11-1 \rangle$)
move on the {101} plane, generating $\langle 100 \rangle$
junction dislocations, a common BCC work hardening mechanism
\cite{DOMINGUEZGUTIERREZ202238}.

We present results for [001] and [011] orientations in 
SM, for single element materials at [001] orientation
(Fig. \ref{fig:WMOVstructure}a), two prismatic dislocation
loops propagate within the Mo matrix, with Burgers vectors
of 1/2$\langle111\rangle$ and symmetrical ones of
1/2$\langle1-11\rangle$. In contrast, W and V samples 
exhibit shear loops developing toward \{111\} slip planes. This consistency aligns with
Fig. \ref{fig:single_samples}, where Mo maintains consistent
p/$E_Y$ values beyond 0.1$a(h)/R_i$, while others experience a
contact pressure drop.
W--Mo and W--V alloys also present these half loops, 
indicating the presence of screw dislocations. Interestingly, W and V samples
also show twinning plane formation beneath the indenter tip, 
by the BDA method \cite{PhysRevMaterials.7.043603}. 
This suggests twinning deformation involvement, supplementing 
plastic response under nanoindentation. 
For [110] orientation, W, Mo, 
and V matrices exhibit lasso-like mechanisms, with screw 
dislocations evolving Burgers vectors of 1/2$\langle1-1-1\rangle$ 
and 1/2$\langle111\rangle$, forming dislocation loops. 
W forms perfect loops, while Mo and V generate distorted
ones due to material--specific mechanical properties
and Peierls barrier effects~\cite{GRIGOREV2023118734}. 
In alloys, V's influence on W matrices surpasses Mo's. 
In W--Mo, two dislocation loops nucleate. However, 
with 50\% V atoms in W, screw dislocation propagation 
on \{111\} plane is significantly absent.

\begin{figure}[t!]
    \centering
    \includegraphics[width=0.48\textwidth]{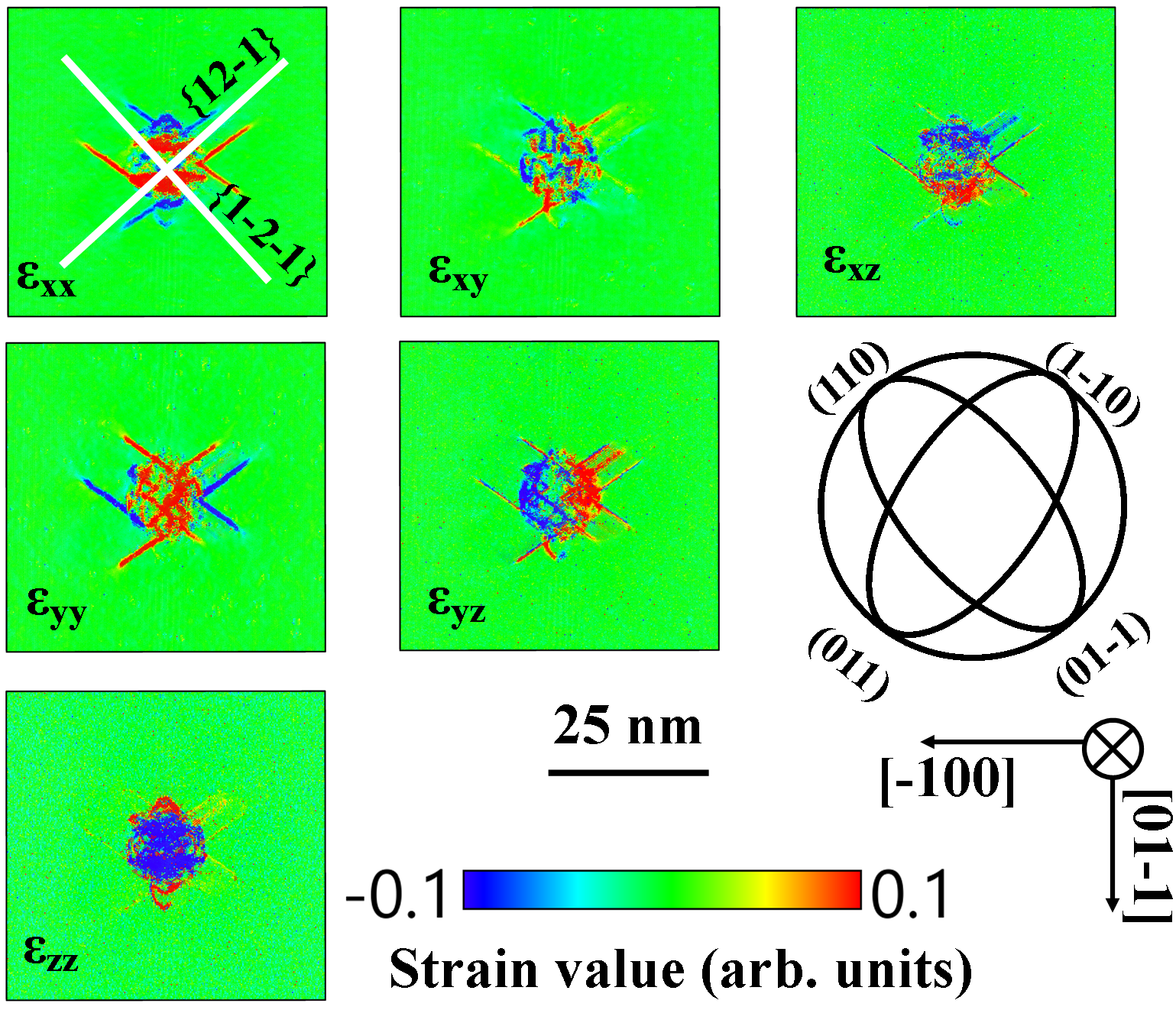}
    \caption{Strain field mapping around the indenter tip 
    at the maximum depth for [001]W noticing a good 
    qualitative agreement with experimental results reported 
    by Yu et al. \cite{YU2020135} following \{112\} 
    plane families and kikuchi-wise pattern with (011) directions.}
    \label{fig:SingleStrainMapping}
\end{figure}

During nanoindentation loading, the indenter tip forms
pile--ups and slip traces around it 
\cite{KURPASKA2022110639,PATHAK20151}, offering insight
into dislocation propagation \cite{YU2020135}. To explore this,
we calculate the strain tensor at maximum indentation depth
using the elastic Eulerian--Almansi finite strain tensor 
\cite{Stukowski_2012} that describes local elastic 
deformation.
Fig. \ref{fig:SingleStrainMapping} displays strain field
maps for [011] orientation of W in good agreement with
Yu et al.'s experimental results \cite{YU2020135},
aligning High--resolution EBSD data.
Positive max strain ($\epsilon_{xx}$) concentrates under the 
indenter, negative strain appears between \{112\} planes. 
$\epsilon_{xy}$ displays four--fold symmetry, $\epsilon_{xz}$
and $\epsilon_{yz}$ exhibit positive and negative poles
aligned with \{110\} family planes as shown by the 
kikuchi--wise patter \cite{YU2020135}. 
In SM, we report that Mo and V exhibit similar strain 
patterns but with differing magnitudes due to lattice 
parameters and elasticity. V emphasizes strain 
accumulation around \{110\} and \{112\} planes in 
$\epsilon_{xx}$ and $\epsilon_{yy}$. 
These strain maps unveil localized
deformation patterns and crystallographic strain
dependence in the studied materials.
In SM, we illustrate strain distribution for W--Mo and 
W--V alloys, respectively.
W--Mo alloy exhibits notable strain accumulation beneath the
indenter tip along the \{110\} plane, largely in 
$\epsilon_{xx,y,z}$. This arises due to Mo's presence in
the W matrix, influencing strain and causing dislocation
buildup under the tip. The strain mapping depicts
dislocation evolution in this direction.
Conversely, strain mapping of the W--V alloy shows
localized positive and negative strain under the tip,
with no noticeable strain propagation along
crystallographic planes. This indicates minimal dislocation 
propagation in the alloy, consistent with Fig. 
\ref{fig:WMOVstructure}. Understanding strain distribution
and its crystallographic dependency is key to comprehending
mechanical response and deformation mechanisms in these
binary alloys.

\begin{figure}[t!]
    \centering
    \includegraphics[width=0.48\textwidth]{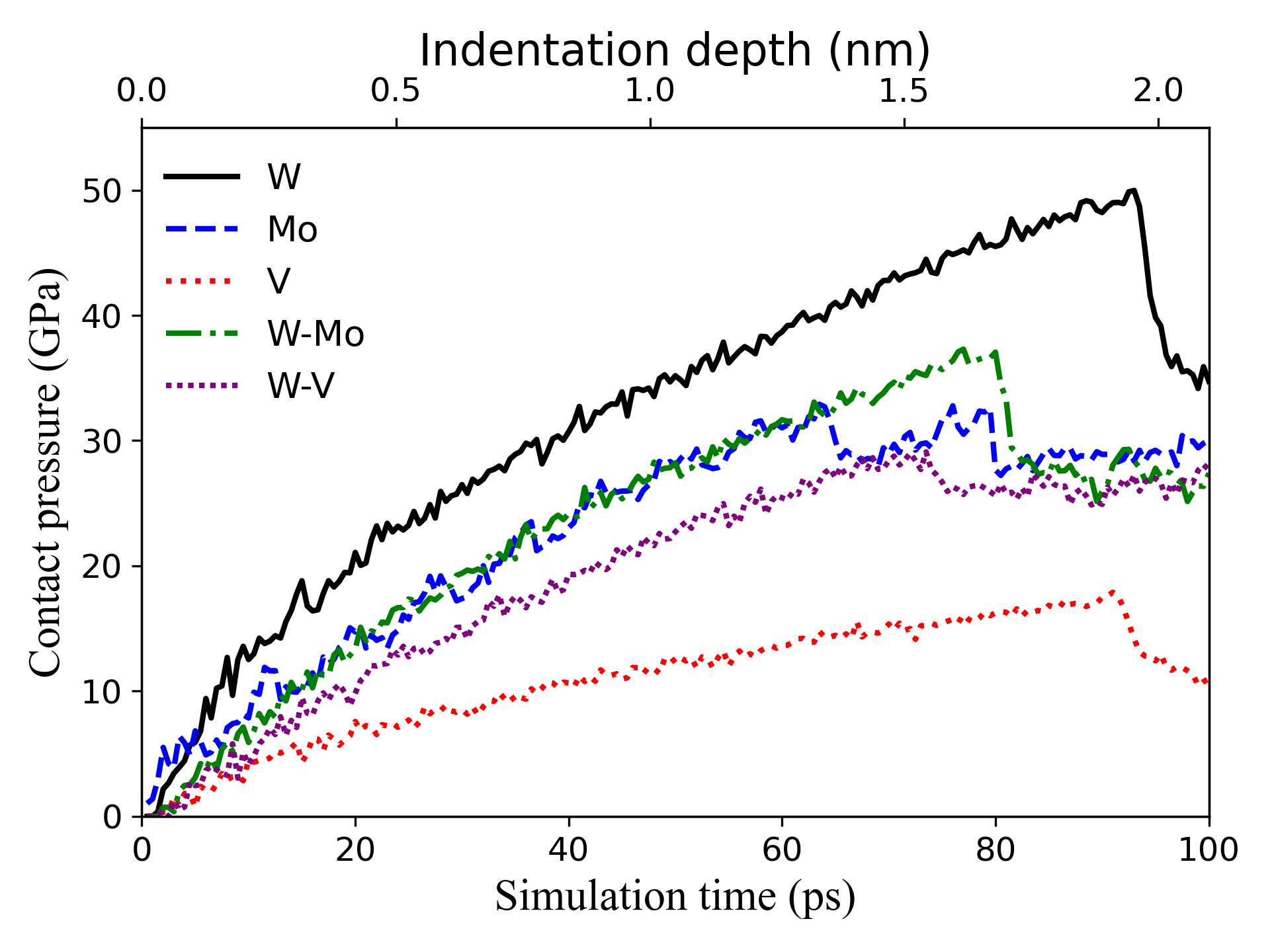}
    \caption{Average contact pressure evolution as a function of 
    simulation time and indentation depth for single element W, Mo, 
    and V matrices at [001] crystal orientation.}
    \label{fig:contacPressure}
\end{figure}

Strain-rate effects are also significant: We perform MD simulations under various nanoindentation strain 
rates at 5 m/s, while maintaining the same numerical setup
as the 20 m/s case. Fig. \ref{fig:contacPressure} present
the results for the average contact pressure, 
$p_m = P_{\rm ave}/\pi a(h)$, for the single--element samples
W, Mo, and V, as well as the binary alloys W--Mo
and W--V for the [001] orientation, 
respectively \cite{ALCALA20102714}.
We observed a sudden drop in the critical contact pressure,
akin to what's observed in compression mechanical 
tests~\cite{papanikolaou2012quasi,papanikolaou2017obstacles,xu2021molecular,song2019discrete,song2019universality}, 
linked to the pop--in event during loading. 
Furthermore, the yield point, signifying the initiation
of plastic deformation, varied among materials. 
In MD simulations, the sequence of yield points is 
W, Mo, and V indicating that tungsten (W) demonstrated
the highest resistance to plastic deformation.
These variations in yield points is crystallographically
dependent and can be attributed to the inherent hardness
of the materials. 
Additionally, the contact pressure decreases with decreasing 
nanoindentation strain rate, primarily observed in the
binary alloys, consistent with experimental findings
\cite{wenyi}. This effect is due to the slower indenter
tip penetration, allowing more time for dislocation
evolution within the sample and shock absorption between
the tip and the surface.
In W--Mo alloys, the critical pressure value falls between
that of Mo and W for both the [011] and [001] orientations,
as expected.
In W--V alloys, regardless of crystal orientation, the critical 
contact pressure value falls between that of V and W.
Results for [001] and [011] orientations are shown in SM.

In our prior study \cite{PhysRevMaterials.7.043603}, we
approximated the onset of dislocation nucleation with
$\sim0.45-0.5z/a$, where $z$ is indentation depth and $a$
is contact area. Therefore, for estimating the maximum
shear stress using contact pressure calculations, we assume an 
isotropic Hertzian contact model \cite{LI20111147}. The maximum 
shear stress beneath the indenter tip ($\tau_{\rm Max}$) 
is computed as $\tau_{\rm Max} = 0.31 p_0$, where $p_0$ is the 
maximum contact pressure under the indenter tip. $p_0$ is derived 
from the mean contact pressure ($p_m$) as $p_0 = 1.5 \max(p_m)$ 
\cite{LI20111147,XIONG2016214}.
The obtained maximum shear stress under the indenter tip
for single elements (W, Mo, V) and binary alloys (W--Mo,
W--V) across various crystal orientations ([001], [011],
and [111]) is shown in Table \ref{tab:MaxShearNano}. 
The crystallographic orientation significantly influences
both the contact pressure yield point and maximum shear
stress values. To determine the maximum shear stress, 
we used the shear modulus (G) computed using the 
EAM--FS potentials at 300K.
The $\tau_{\rm Max}$ beneath the indenter tip agrees 
well with $\tau_{\rm theor}$ for the [001] orientations 
across all samples. While the results for single 
element Mo matrices deviated from theoretical expectations 
being a factor to slow down 
dislocation evolution, indicating their influence on material 
behavior.
Furthermore, the W--V alloy displayed heightened ductility 
for the [111] crystal orientation. This was evident in the
contact pressure analysis, where the indenter tip needed to 
penetrate deeper into the sample to reach the yield point.

\begin{table}[t!]
    \centering
    \begin{tabular}{crrr|rr|r}
    \hline \hline
             & W   & Mo & V & W--Mo & W--V & Ind. rate \\
    \hline 
      [001]  & \textbf{22.50} & \textbf{14.34} & \textbf{7.52} & \textbf{18.58} & \textbf{15.31}  & 20 m/s\\
             & 22.95 & 14.85 & 8.01 & 19.26 & 14.85 & 5 m/s\\
    \hline
      [011]  & \textbf{18.22} & \textbf{13.61} & \textbf{6.52} & \textbf{13.77} &\textbf{13.99} & 20 m/s\\
             & 17.32          & 11.11  & 5.4 & 13.86 & 13.51 
             & 5 m/s\\
    \hline
    [111]   &  \textbf{19.57} & \textbf{13.05} & \textbf{6.06} & \textbf{15.12} & - & 20 m/s \\
            &  18.55 & 12.73 & 6.98 & 14.53 & - & 5 m/s\\
    \hline
    G       &  136  & 122 & 43 &  135 & 84 & 20 m/s\\
    Theor.  &  21.65 & 19.45 & 6.84 & 21.56 & 13.37 & 5 m/s  \\
    \hline
    \end{tabular}
    \caption{Maximum shear stress, $\tau_{\rm max} = 0.465$max$(p_m)$ \cite{LI20111147}, under the indenter tip for single elements 
     (W, Mo, V) and binary alloys (WMo, WV) at different crystal orientations 
     ([001], [011], [111]) for a strain rate of 20 and 
     5 m/s. Additionally, the theoretical shear stress 
     ($\tau_{\rm theor} = G/(2\pi)$ ) is provided, where $G$ represents the shear modulus.}
    \label{tab:MaxShearNano}
\end{table}



Summarizing, through extensive MD simulations, 
we explored defects and dislocation mechanisms 
using a BCC defect analysis tool. Our approach revealed 
\{112\} plane twinning emergence in the alloys and accurately 
depicted plastic deformation processes in pure W, Mo, and V 
samples which is related to the strain mapping patterning 
at the maximum indentation depth. 
Intriguingly, Mo and V addition in the W matrix 
heightened critical load and maximum shear stress during the 
pop--in event, signifying the shift from elastic to plastic 
deformation. 

We acknowledge support from the European Union Horizon 2020 
research
and innovation program under grant agreement no. 857470,
from the European Regional Development Fund via 
the Foundation for Polish Science International 
Research Agenda PLUS program grant No. MAB PLUS/ 2018/8, 
and  INNUMAT project (Grant Agreement No. 101061241). 
We acknowledge the computational resources 
 provided by the High Performance Cluster at the National Centre 
 for Nuclear Research and 
 the Interdisciplinary Centre for Mathematical and
 Computational Modelling (ICM) University of Warsaw under 
 computational allocation no g91--1427.

\section*{DATA AVAILABILITY}
The data that support the findings of this study are available
from the corresponding author upon reasonable request.
\section*{REFERENCES}
\bibliography{bibliography,biblio}
\bibliographystyle{iopart-num}

\end{document}